\documentclass[useAMS,usenatbib]{mn2e}
\usepackage[dvips]{graphicx}
\title[Desorption From Interstellar Ices]{Desorption From Interstellar Ices}
\author[J. F. Roberts, J. M. C. Rawlings, S. Viti and D. A. Williams]{J. F. 
Roberts\thanks{E-mail: jfr@star.ucl.ac.uk}, J. M. C. Rawlings, S. Viti and D.
A. Williams\\
Department of Physics \& Astronomy, University College London, Gower Street, 
London WC1E 6BT}
\begin{document}

\pagerange{\pageref{firstpage}--\pageref{lastpage}} \pubyear{2007}

\maketitle

\label{firstpage}

\begin{abstract}

The desorption of molecular species from ice mantles back into the gas phase
in molecular clouds results from a variety of very poorly understood 
processes. We have investigated three mechanisms;
desorption resulting from H$_2$ formation on grains, direct cosmic ray heating 
and cosmic ray induced photodesorption. 
Whilst qualitative differences exist between these processes (essentially 
deriving from the assumptions concerning the species-selectivity of the
desorption and the assumed threshold adsorption energies, $E_\mathrm{t}$) all
three processes are found to be potentially very significant in dark 
cloud conditions. 
It is therefore important that all three mechanisms should be considered in
studies of molecular clouds in which freeze-out and desorption are believed to
be important.

Employing a chemical model of a typical static molecular core and 
using likely estimates for the quantum yields of the three processes we find
that desorption by H$_2$ formation probably dominates over the other two 
mechanisms.
However, the physics of the desorption processes and the nature of the dust
grains and ice mantles are very poorly constrained. We therefore conclude that
the best approach is to set empirical constraints on the desorption, based on
observed molecular depletions - rather than try to establish the desorption
efficiencies from purely theoretical considerations.
Applying this method to one such object (L1689B) yields upper limits to the
desorption efficiencies that are consistent with our understanding of 
these mechanisms.

\end{abstract}


\section{Introduction}

In cold dark interstellar clouds, heavy molecules accumulate onto dust grains, 
forming icy mantles on their surfaces.
This process, known as freeze-out, occurs on a timescale of a few times
$10^9 n_\mathrm{H}^{-1}$ years (where $n_\mathrm{H} = n(\mathrm{H}) + 
2 n(\mathrm{H}_2)$ is the total hydrogen
nucleon number density) in the absence of desorption.  This is much less 
than the expected lifetime of a typical molecular cloud
so, if freeze-out was unlimited, we would expect the majority of observations 
to show no evidence for heavy gas phase species.  
However, observations of molecules such as CO in dark clouds, for example 
L1689B, TMC1-CP and L134N  \citep{lee,b27,b30}, indicate that desorption 
processes must be operating for mantle growth to be limited. 

Although this conclusion has been accepted for over 20 years, it is still 
not fully 
understood how this desorption occurs.  Many possible mechanisms have been 
proposed, most of which require impulsive heating of grains which can be caused
by (a) direct impact of cosmic rays \citep{b17,b16,b18}, (b) X-rays 
\citep{b18}, (c) ultra-violet photons induced by cosmic rays (cosmic ray 
photodesorption) \citep{b16,b7}, or (d) exothermic reactions occurring on the 
grain surface \citep{b2,b14}, in particular the formation of molecular 
hydrogen \citep{b32,b8}.  Molecules can either be desorbed by classical 
evaporation \citep{b18} or by chemical explosions \citep{b26,b25,b18}. These
chemical explosions can only occur if the mantle has previously been irradiated
with ultra-violet radiation which creates radicals. If the grain temperature 
is raised to $\sim 27$~K, the radicals become mobile and then undergo explosive
reactions capable of expelling the entire grain mantle.  We will not consider 
chemical explosions in this paper because in dark clouds with $A_V > 5$ 
magnitudes it is unlikely that the grain mantles will have received sufficient
UV irradiation \citep{b18}, and any radicals that do form are likely to be 
hydrogenated due to the high abundance of hydrogen atoms present \citep{WM98}.

Recent chemical models of dark clouds tend to include only desorption via 
thermal evaporation (which is negligible for dark clouds with temperatures of 
10~K) and/or direct heating by cosmic rays, using the formulation given by 
\citet{b17} (hereafter HH93) \citep{ruf,rob}. 
This paper, therefore, aims to test the assumption that desorption via direct 
cosmic ray heating is the only effective non-thermal desorption mechanism
relevant to dark molecular cores. In this study we adopt the model of a 
molecular cloud as being composed of an ensemble of dense (dark) cores in a
more diffuse background \citep{b13}.
Ices are probably only present in the cores.
We set out to test the desorption efficiencies, 
by including in an existing model
of dark cloud chemistry three desorption mechanisms (desorption resulting from 
H$_2$ formation on grains, direct cosmic ray heating and cosmic ray 
photodesorption) that have been formulated in the literature, to find their 
relative importance in dark cores.  
We believe these three mechanisms are likely to be the most important in this
situation, in the absence of any nearby X-ray sources in the molecular cloud.

We note that there already exist several other studies which model these three 
desorption mechanisms, for example \citet{b33} (hereafter WRW94) 
performed a study very close to our 
own.  However, our work differs to that of WRW94 because we try to take a 
simpler approach; by modelling a static cloud rather than a collapsing one we 
can easily see the effects of adding desorption to our chemical model.  We 
also look further into the assumptions made about the ability of each 
desorption mechanism to desorb molecules with higher adsorption energies, such 
as H$_2$O and NH$_3$ (Section~\ref{sec:threshold}).  Most importantly, 
we attempt to 
constrain the efficiency of each mechanism by comparing our results to 
observations of CO depletion in star-forming regions.

The paper is structured as follows:  In Section~\ref{sec:des} we give a brief
summary of these desorption mechanisms
and in Section~\ref{sec:model} we describe the model.
The results are given in Section~\ref{sec:results}, and are discussed by
comparing them to existing models in Section~\ref{sec:comp_models}
and observations in Section~\ref{sec:comp_obsns}.
Concluding remarks are given in Section~\ref{sec:conclusions}.

\section{Desorption mechanisms}
\label{sec:des}

\subsection{Desorption resulting from H$_2$ formation}
\label{sec:H2}

It has long been suggested that the energy released from exothermic reactions
on grain surfaces can release energy capable of desorbing mantle species
\citep{b2}.
In particular, laboratory experiments on graphite substrates suggest that a
non-negligible fraction, perhaps up to 40\% \citep{h2exp}, (F. Islam,
private communication) of the
$\sim$~4.5~eV released in the surface formation of molecular hydrogen is
transfered to the grain surface. This leads to local heating which may
thermally desorb weakly bound mantle species \citep{b8},
although the extent to which the temperature rises transiently is not well
determined, and depends on the local conductivity. In amorphous materials,
the temperature rise may be relatively large.

Such a process may be \textit{selective}, depending on the temperature
achieved as a result of the energy deposition. In previous work
\citep{b8,b32,b33},
the conservative assumption was made that only the most volatile species 
(such as CO, N$_2$, NO, O$_2$, C$_2$ and CH$_4$) would be desorbed during the
transient heating, corresponding to a threshold adsorption energy of
$E_\mathrm{t}=1210$\,K. Initially, we shall adopt the same threshold and range
of volatile species, but we shall also consider the effects of variations 
from that value in Section~\ref{sec:threshold}.

The rate of desorption by this process for species $i$ is given by:
\begin{equation}
\begin{array}{ll}
R_{\mathrm{hf}}=\varepsilon R_{\mathrm{H}_2} M_\mathrm{s}(i,t)& 
\mbox{ cm$^{-3}$ s$^{-1}$}
\end{array}
\end{equation}
where $R_{\mathrm{H}_2}$ is the rate of H$_2$ formation on grains.
$R_{\mathrm{H}_2}$ is proportional to the available grain surface area
and the sticking/reaction probability to form H$_2$ and is
empirically constrained (at T=10~K) to be given by:
\begin{equation}
\begin{array}{ll}
R_{\mathrm{H}_2}=3.16 \times 10^{-17} n(\mathrm{H}) n_\mathrm{H} &
\mbox{cm$^{-3}$ s$^{-1}$}
\end{array}
\end{equation}
where $n(\mathrm{H})$ is the number density of atomic hydrogen and 
$n_\mathrm{H}$ is the total hydrogen nucleon density.
$M_\mathrm{s}(i,t)$ is the fraction of the mantle consisting of species $i$,
calculated self-consistently as a function of time.  
$\varepsilon$ is an efficiency parameter such that $\varepsilon 
M_\mathrm{s}(i,t)$ gives the number of molecules of species $i$ desorbing 
per H$_2$ molecule formed.  
The efficiency of this process is uncertain, but we can make a rough upper
estimate as follows:
If we assume that 40\% ($\sim 2$~eV) of the energy released during each
H$_2$ formation is transfered to the grain, and given that the adsorption 
energies of the species we consider are $\sim 1000$~K ($\sim 8.2
\times 10^{-2}$~eV), we would expect that a maximum of $\sim 20$ molecules
could be desorbed every time an H$_2$ molecule is formed.
It is likely, however, that this process is less efficient, because such a high
value for $\varepsilon$ would prevent the build-up of any mantle material until
very late times \citep{WM98}.  We therefore run our model with values of
$\varepsilon$ ranging from 0.01 to 1.0, and present results for $\varepsilon
=0.01$ and 0.1.

\subsection{Desorption by direct cosmic ray heating}

In this paper, the rate of desorption by direct cosmic ray heating is 
calculated by simply considering the number of molecules capable of being 
desorbed per cosmic ray impact. 
This rate is different to that adopted in other models (see the discussion in
Section~\ref{sec:HH93}).

As with desorption via H$_2$ formation, this process is believed to be 
selective \citep{b33}, so only the volatile species (CO, etc.) are expected to
be desorbed, at a rate given by: 
\begin{equation}
\begin{array}{ll}
R_{\mathrm{cr}}=F_{\mathrm{cr}}\langle \pi a_\mathrm{g}^2 n_\mathrm{g} \rangle 
\phi M_\mathrm{s}(i,t)& \mbox{cm$^{-3}$ s$^{-1}$.}\end{array}
\label{eqn:cr}
\end{equation}
where $F_{\mathrm{cr}}$ is the flux 
of cosmic rays, $\phi$ is an efficiency parameter such that $\phi 
M_\mathrm{s}(i,t)$ is the number of molecules released per cosmic ray impact,
$a_\mathrm{g}$ is the grain radius and $n_\mathrm{g}$ is the number density of 
grains.  We have used a value for the total grain surface area per cm$^3$, 
$\langle \pi a_\mathrm{g}^2 n_\mathrm{g} \rangle$, of $2.4\times10^{-22}
n_\mathrm{H}$~cm$^{-1}$, consistent with \citet{b24} if we use a value for the 
depletion coefficient of $\sim 1.0$.

\citet{b18} showed that the iron nuclei component of cosmic rays should be the
most effective in heating the grain, and therefore we adopt for the value for 
$F_{\mathrm{cr}}$ of $2.06 \times 10^{-3}$ cm$^{-2}$ s$^{-1}$ which is an
estimate of the
iron fraction of the canonical cosmic ray flux. It is based on the proton 
cosmic ray flux in the interstellar medium given by \citet{b22}, and the iron 
to proton ratio of $1.6 \times 10^{-4}$ (estimated by \citet{b18}).

An estimate for $\phi$ can be taken from \citet{b18} who worked out the 
evaporation rate of CO per grain resulting from both whole-grain and spot-grain 
heating by cosmic rays.
For spot-grain heating, they calculated that $6 \times 10^4$ CO molecules
would be released per cosmic ray impact, independent of grain size.  For whole
grain heating, however, the result was found to strongly depend on grain size,
dominating spot-grain heating for $a_\mathrm{g} \le 2500$~\AA.
We have tested this mechanism with
values of $\phi$ ranging from $10^2$ to $10^6$, but we only present the results
for $\phi=10^5$ to be consistent with the lower estimate predicted by
spot-grain heating.
Since the volatile species make up approximately 40\% of the grain mantle,
using $\phi=10^5$ implies that the number of molecules released per cosmic ray
impact is $\phi M_\mathrm{s}(i) \approx 4 \times 10^4$.

\subsection{Cosmic ray induced photodesorption}

As cosmic rays travel through a molecular core, they ionise and excite the
absorbing gas. \citet{b23} investigated the excitation of the Lyman and Werner 
systems of the hydrogen molecule, caused by collisions with either primary 
cosmic ray particles or secondary electrons released by cosmic ray ionisation.  
They found that emissions resulting from these excitations may be capable of 
maintaining a chemically significant level of UV photon flux in the interiors 
of dark clouds, where the interstellar UV radiation is heavily extinguished 
($A_V \ge 5$). 

When a UV photon impinges upon the mantle of a grain, 
it is most likely to be absorbed by H$_2$O which may then be dissociated into H 
and OH \citep{b16}.  The energy of the dissociation products cause local 
heating, capable of desorbing nearby molecules.  This process is non-selective, 
and proceeds at a rate given by:
\begin{equation}
\begin{array}{ll}
R_{\mathrm{crpd}}=F_{\mathrm{P}}\langle \pi a_\mathrm{g}^2 n_\mathrm{g} \rangle 
Y  M_\mathrm{s}(i,t)& \mbox{cm$^{-3}$ s$^{-1}$.}\end{array}
\end{equation}
where $F_\mathrm{P}$ is the photon flux and $Y$ is the yield per photon. 
\citet{b23} estimated $F_\mathrm{P}$ to be just 1350~cm$^{-2}$ s$^{-1}$, but
here we use a value of $4875$~cm$^{-2}$~s$^{-1}$ taken from equation (21) of 
\citet{b3} \footnote{This is consistent with using a cosmic ray ionisation 
rate, $\zeta$, of $1.3 \times 10^{-17}$~s$^{-1}$.}.  
This value is much higher than the estimate of \citet{b23} because it
explicitly includes the contribution to the excitation rate of H$_2$ by cosmic 
ray protons, not just the secondary electrons.
We present the results for $Y=0.1$, as estimated by \citet{b16}, but as this
value is uncertain we also ran the model with values of $Y$ ranging from 
$1 \times 10^{-3}$ to 100.

Obviously, the relative importance of these mechanisms depends upon the values 
used for $\varepsilon$, $\phi$ and $Y$, which can be treated as free 
parameters.  In this paper, we present the results for the values stated above, 
and in Section~\ref{sec:comp_obsns} we derive upper limits for these parameters 
based 
on observations.

\section{The Model}
\label{sec:model}

In order to see clearly the relative importance of the various desorption 
mechanisms for the interstellar chemistry of molecular cores, we 
have used a simple one-point model of a static, isothermal dark cloud, at a
density of $n_\mathrm{H}=10^5$~cm$^{-3}$ and a temperature of 10~K. The core 
has a visual extinction, $A_V$, of 10 magnitudes, and a radius of $\sim 
0.05$~pc, implying a mass of $1$~M$_\odot$.

\subsection{The chemical model}
\label{sec:chemod}

The model, adapted from \citet{b29}, includes gas-phase reactions, freeze-out, 
surface reactions and desorption, with 127 gas phase and 40 mantle species.  
The gas phase reaction network consists of 1741 reactions from the UMIST Rate99 
database \citep{b20}.
Photoreactions are included, which take into account both the external 
interstellar radiation field and the internal cosmic ray induced UV field.  
Both direct and indirect ionisation by cosmic rays are also included, using a 
cosmic ray ionisation rate, $\zeta$, of $1.3 \times 10^{-17}$~s$^{-1}$.
The hydrogen atom abundance (that drives the H$_2$ formation desorption
mechanism) is determined self-consistently in the chemical network.
The species and reactions are taken from the model as in \citet{b29}, which 
describes the rich chemistry of hot cores.
The number of species and elements should therefore be more than sufficient to
describe the low mass case in this paper.
Freeze-out and grain chemistry are
described in Section~\ref{sec:FO}.

The initial conditions are atomic, apart from carbon which is all singly
ionised, and half the hydrogen nuclei are in the form of H$_2$. The initial
elemental abundances are taken from \citet{sof}.

\subsection{Freeze-out and grain chemistry}
\label{sec:FO}

The rate of accretion onto grains of species $i$ (in cm$^{-3}$s$^{-1}$)
is given by \citep{b24}:
\begin{equation}
\frac{dn(i)}{dt}= - 4.57 \times 10^4 \langle  \pi a_\mathrm{g}^2 n_\mathrm{g}
\rangle T^{1/2} C S(i) m(i)^{-1/2} n(i)
\end{equation}
where $T$ is the gas temperature,
$m(i)$ is the mass of species $i$ in atomic mass units,
$n(i)$ is the number density of species $i$, 
 $\langle  \pi a_\mathrm{g}^2 n_\mathrm{g} \rangle$ has the same definition
as in Section~\ref{sec:des},  
and $S(i)$ is the sticking coefficient of species $(i)$.
Following WRW94, we have used $S(i) = 0.3$ for all species.  

$C$ is a factor which takes into account electrostatic effects:
\begin{equation}
\begin{array}{lll}
C & =1 &\mbox{for neutral species}\\
 & = 1+(16.71 \times 10^{-4}/a_\mathrm{g} T) & \mbox{for singly charged}\\ 
&&\mbox{positive ions.}
\end{array}
\end{equation} 

The only surface reactions that occur are hydrogenation of certain unsaturated 
species, and dissociative recombination of molecular ions.  These reactions are
assumed to occur instantly and the products remain on the grain surface until
they are desorbed by one of the processes named above.  H$_2$ is an exception
in that it is assumed to desorb immediately because of the high exothermicity of
the reaction, as is indicated by experimental studies \citep{h2exp}.

\section{Results}
\label{sec:results}

By comparing the rate coefficients given for each of the desorption mechanisms,
if we use $\varepsilon=0.01$, $\phi=10^5$ and $Y=0.1$ at a density of
$n_\mathrm{H}=10^5$~cm$^{-3}$, desorption via H$_2$ formation will dominate if
the equilibrium atomic hydrogen density is greater than $\sim 0.4$ cm$^{-3}$. 
 If $\varepsilon$ is as high as $0.1$ then this mechanism will always be
dominant.

\begin{figure*}
\includegraphics{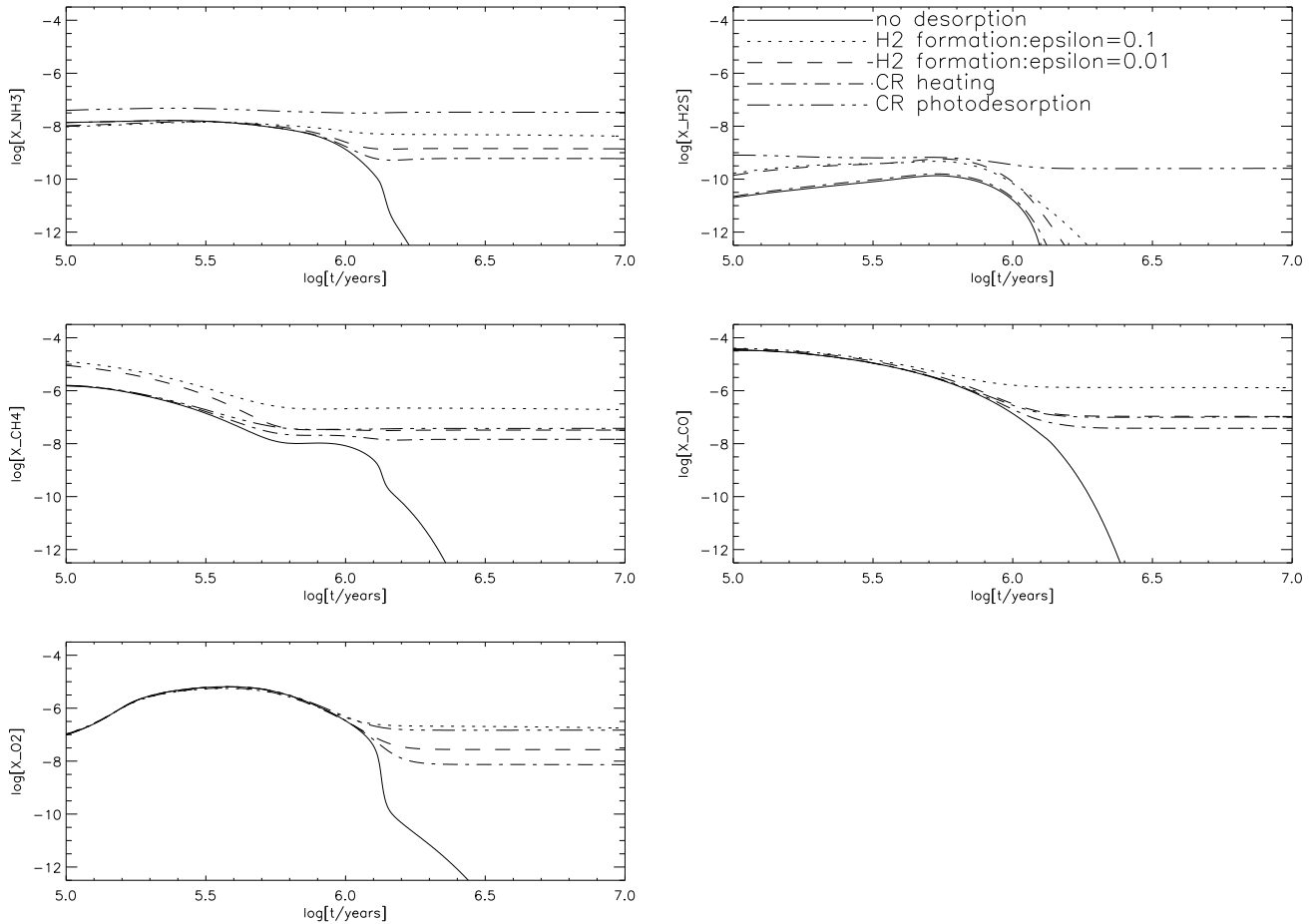}
\caption{The time evolution of the molecules NH$_3$, H$_2$S, CH$_4$, CO and
O$_2$ in a static cloud of density $n_\mathrm{H}= 10^5$ cm$^{-3}$, $T~=~10$~K,
at a point with $A_V = 10$ mag.  The abundances of species relative to hydrogen
are shown as a function of time in years.  The different curves compare the
evolution of the species with no desorption, desorption via H$_2$ formation
with $\varepsilon = 0.1$ and 0.01, desorption via direct cosmic ray heating and
cosmic ray photodesorption (see key).}
\label{fig:mech1}
\end{figure*}

\begin{figure*}
\includegraphics{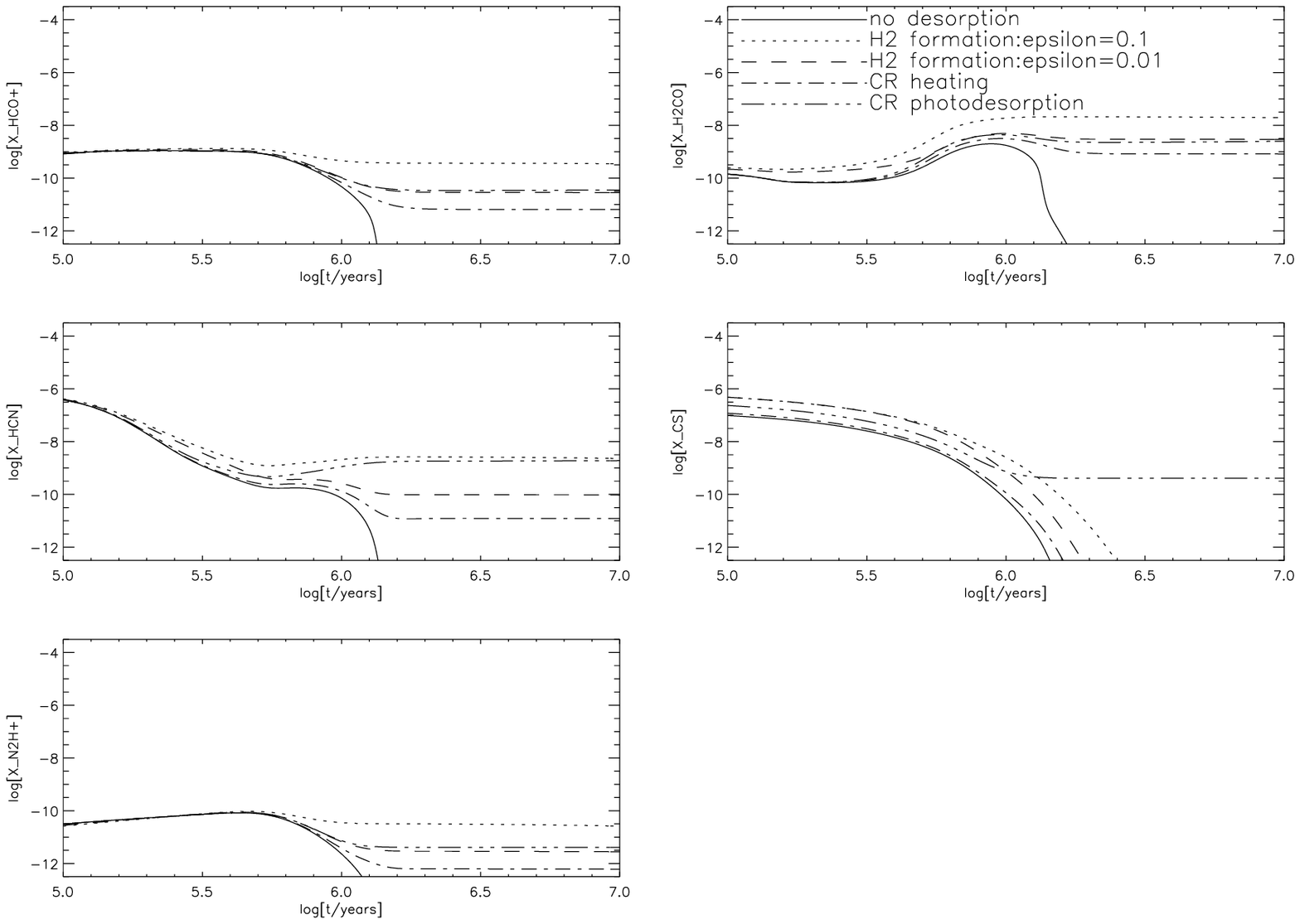}
\caption{As in Figure~\ref{fig:mech1} but for the molecules HCO$^+$, H$_2$CO,
HCN, CS and N$_2$H$^+$.}
\label{fig:mech2}
\end{figure*}

Figures~\ref{fig:mech1} and~\ref{fig:mech2} compare the effect of the different
desorption mechanisms in a static cloud at a density of $10^5$~cm$^{-3}$ for
selected observable molecules.  In general, desorption has the most significant
effect on gas phase abundances after approximately $10^6$ years, which is when
the molecules begin to freeze-out.  Each of the desorption mechanisms are
effective enough to be able to compete with freeze-out at late times,
preventing full depletion of species onto grains.
The exceptions are the sulphur bearing species, H$_2$S and CS, whose freeze-out
is only prevented by the non-selective cosmic ray photodesorption.  This
mechanism is also able to enhance abundances of molecules such as NH$_3$ and
H$_2$S at times as early as $10^5$ years.  For the case of NH$_3$ this is
because the NH$_3$ in the mantle builds up relatively quickly (compared to
molecules like CO), so direct desorption by this mechanism can proceed close to
its maximum rate.  However, since the H$_2$ formation and direct cosmic ray
heating mechanisms are not able to desorb NH$_3$ directly, desorption via these
mechanisms can only affect NH$_3$ once enough NO has been desorbed to enhance
the gas phase production (via the reaction $\mathrm{NO} + \mathrm{NH}_3^+ \to
\mathrm{NH}_3 + \mathrm{NO}^+$).

For most of the species shown in Figures~\ref{fig:mech1} and~\ref{fig:mech2}, 
desorption via H$_2$ formation with $\varepsilon=0.1$ is the most effective 
mechanism, being able to produce equilibrium abundances up to an order of 
magnitude greater than the other mechanisms.  However, there are several 
molecules (NH$_3$, H$_2$S and CS), for which cosmic ray induced photodesorption
is more effective, even though desorption via this mechanism proceeds at a rate 
more than ten times slower than  H$_2$ formation with $\varepsilon=0.1$.  This 
is because we have assumed desorption via H$_2$ formation is selective, only 
being able to desorb molecules with adsorption energies less than 1210~K (CO, 
NO, N$_2$, O$_2$, C$_2$ and CH$_4$).  Evidently, the higher gas phase 
abundances of these molecules caused by selective desorption do not have a 
significant effect on the gas phase chemistry of molecules like H$_2$S and CS, 
so their freeze-out is not prevented by selective mechanisms.  

In the following subsections we look at the effectiveness of each mechanism
under different conditions, by varying the density and the initial atomic
hydrogen density.  We also discuss further the effects of selectivity in 
Section~\ref{sec:threshold}.

\subsection{Varying the density} 
When the model was run at a density of $n_\mathrm{H} = 10^6$ cm$^{-3}$,  
it was found that the equilibrium abundances of the desorbed species were 
reduced approximately by a factor of 10 compared to the equivalent runs at 
$10^5$ cm$^{-3}$.  Given that $n(\mathrm{H})$ and $M_\mathrm{s}$ appear to show 
little variation in their equilibrium values at different densities, the rates 
for each desorption mechanism should not depend on density.  However, the rate 
for freeze-out has a direct linear dependence on density, so it is expected 
that the equilibrium abundances of desorbed species should scale roughly as 
$n_\mathrm{H}^{-1}$, which would explain the reduction in equilibrium 
abundances found in the results.

\subsection{Varying the initial atomic hydrogen density}
Since desorption via H$_2$ formation depends on the abundance of atomic 
hydrogen, it would be expected that varying the initial ratio of atomic to 
molecular hydrogen could also influence how efficient this mechanism is.  
However, even when the model was run with all hydrogen in the form of H$_2$ 
initially, after $10^5$ years the abundance of atomic hydrogen had reached the 
same value ($n(\mathrm{H}) \sim 0.4$~cm$^{-3}$) as in the previous runs (which 
had only half the hydrogen nuclei in H$_2$ initially), so there was no 
noticeable difference in the desorption via H$_2$ formation rate. 

\subsection{Varying the threshold adsorption energy for selective mechanisms}
\label{sec:threshold}

So far we have assumed that the selective desorption mechanisms (H$_2$ 
formation and direct cosmic ray heating) are only capable of desorbing 
molecules with adsorption energies  less than or equal to the  value 
$E_\mathrm{t} = 1210$~K.  This number was chosen to be consistent with the 
study by WRW94, who assumed that only CO, N$_2$, C$_2$, 
O$_2$ and NO could be desorbed by these processes. 
CH$_4$ also has an adsorption energy of less than 1210~K, so we have also 
included CH$_4$ as a `volatile' species.  Although \citet{b18} calculated that 
mantles composed {\it purely\/} of refractory ices such as CO$_2$, H$_2$CO, 
HCN, NH$_3$ and H$_2$O are unlikely to be affected by these mechanisms, they 
predicted that if these molecules are mixed with volatile species then spot 
heating processes may be able to raise the temperature of the grains enough to 
desorb these refractory molecules.  Since the H$_2$ formation and direct cosmic 
ray heating mechanisms are both capable of spot-heating, we thought it was
necessary to investigate the effect of varying $E_\mathrm{t}$.

\begin{figure*}
\includegraphics{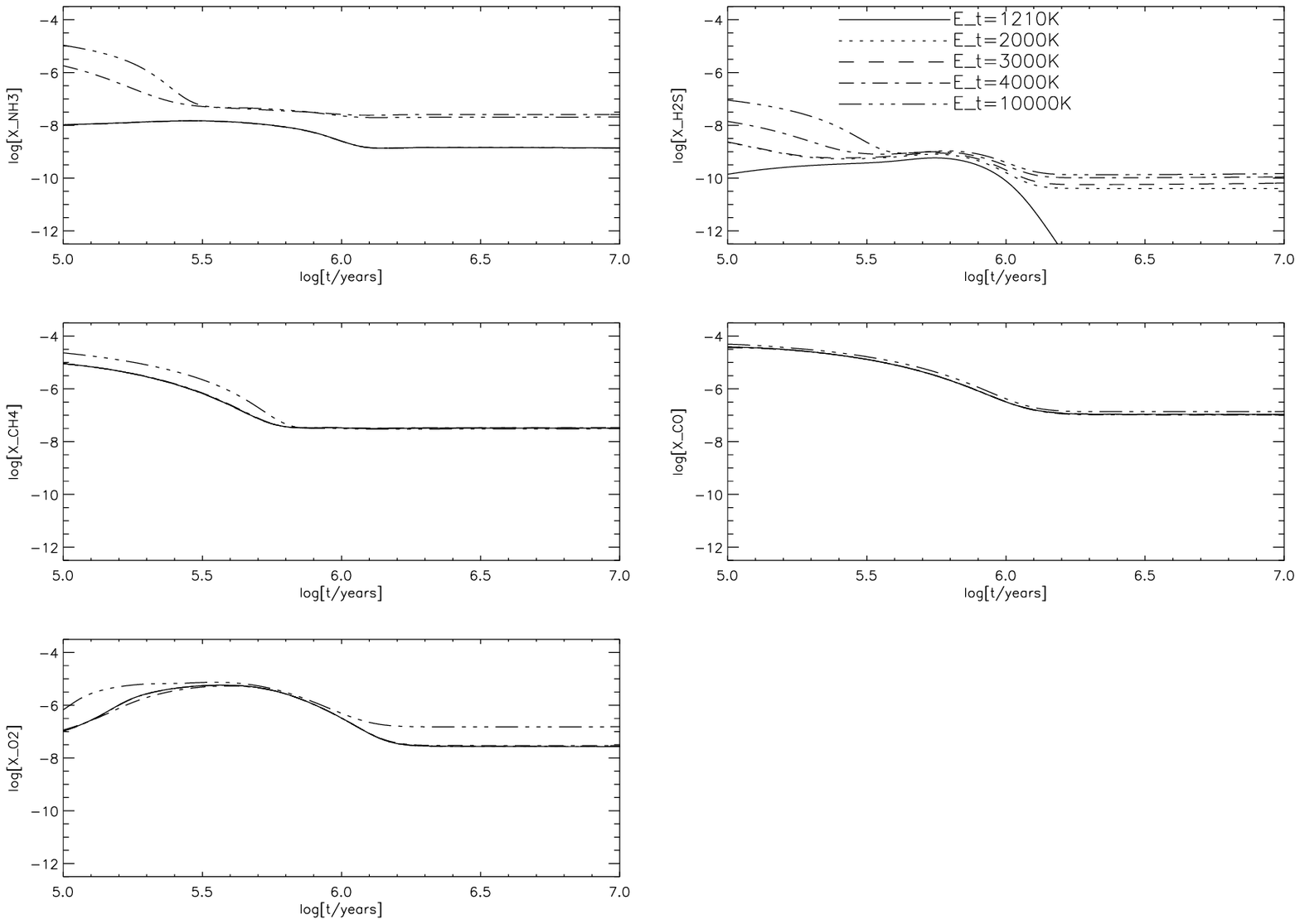}
\caption{The effect of varying $E_\mathrm{t}$ for desorption via H$_2$ 
formation with $\varepsilon~=~0.01 $, in a static cloud of density 
$n_\mathrm{H} = 10^5$ cm$^{-3}$ at 10 K, for the molecules  NH$_3$, H$_2$S, 
CH$_4$, CO and O$_2$}.
\label{fig:ebmax1}
\end{figure*}

\begin{figure*}
\includegraphics{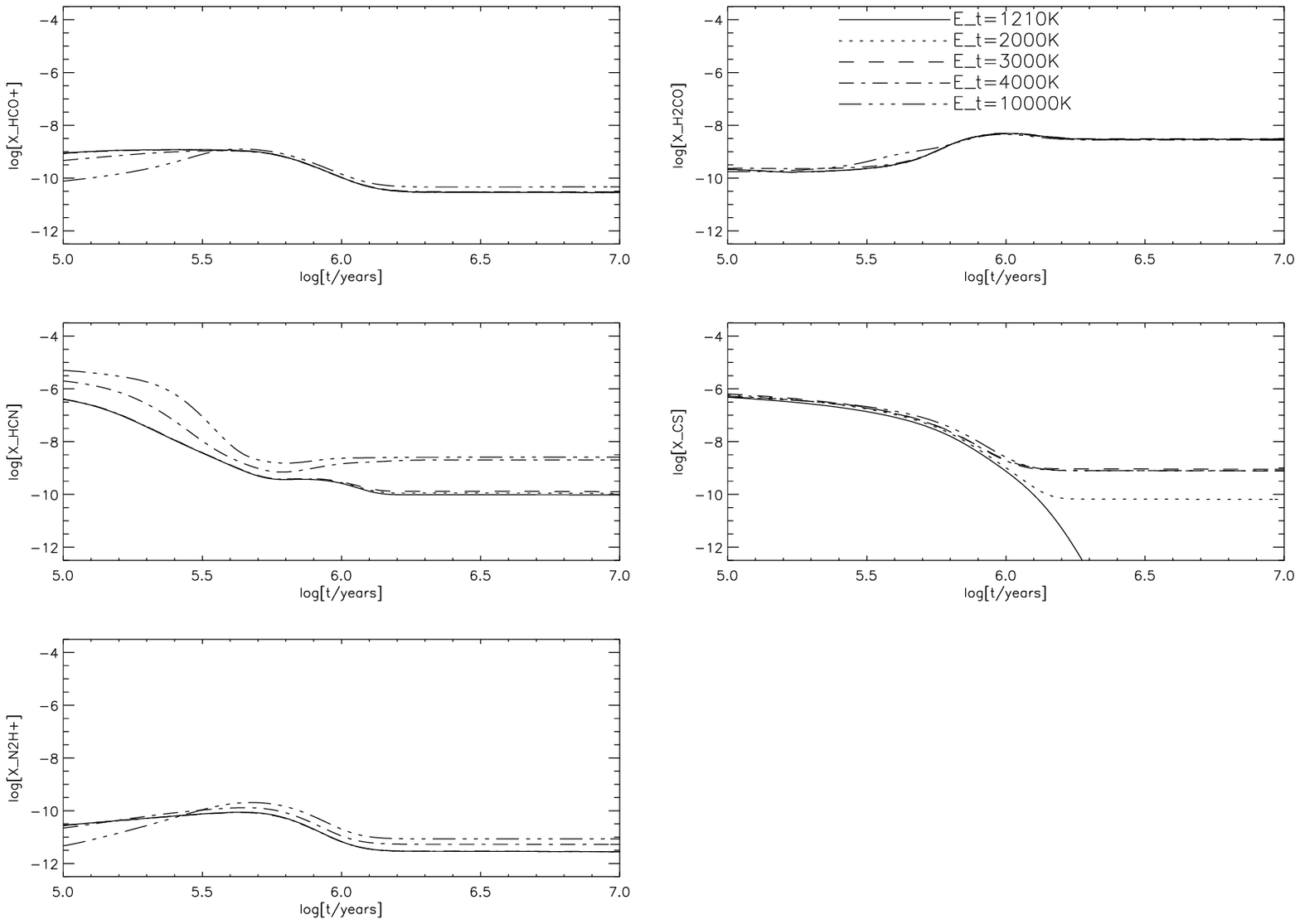}
\caption{As in figure~\ref{fig:ebmax1} but for the molecules  HCO$^+$, H$_2$CO,
HCN, CS and N$_2$H$^+$}.
\label{fig:ebmax2}
\end{figure*}

Figures~\ref{fig:ebmax1} and~\ref{fig:ebmax2} show the effect of varying 
$E_\mathrm{t}$ for the H$_2$ formation mechanism with $\varepsilon = 0.01$.  
The adsorption energies for each mantle species were taken from \citet{b1},
apart from H$_2$O which was taken from \citet{b12}.
Some of the molecules, such as CO, CH$_4$ and H$_2$CO are relatively 
unaffected, even when $E_\mathrm{t}$ is so high (10 000~K) that all molecules 
can be desorbed. In the case of H$_2$CO which has an adsorption energy of
1760~K \citep{b1} this may seem rather surprising. The behaviour can be
explained by the fact that, for the model and parameters that we have
considered, gas-phase formation of H$_2$CO (via CH$_3$+O) dominates over
desorption.
On the other hand, molecules such as NH$_3$ and HCN show differences in 
abundances of more than one order of magnitude when they reach equilibrium.  
These differences can be explained as follows:
\begin{enumerate}
\item NH$_3$:  The equilibrium fractional abundance of NH$_3$ increases from 
$1.4 \times 10^{-9}$ to $2.5 \times 10^{-8}$ when $E_\mathrm{t}$ increases from 
3000~K to 4000~K.  This change is simply because the adsorption energy of 
NH$_3$ is 3080~K, so for $E_\mathrm{t} \geq 4000$~K, NH$_3$ can be desorbed 
directly.  For $E_\mathrm{t} \leq 4000$~K, the main production of NH$_3$ is 
through the reaction NH$_3^+$ + NO $\to$ NO$^+$ + NH$_3$.
\item HCN:  The equilibrium fractional abundance of HCN increases from $9.5 
\times 10^{-11}$ to $2.0 \times 10^{-9}$ when $E_\mathrm{t}$ increases from 
3000~K to 4000~K.  The adsorption energy of HCN is actually greater than 
4000~K, so this is not a consequence of direct HCN desorption.  In fact, it is 
the direct desorption of NH$_3$ that produces the HCN via the reaction 
$\mathrm{NH}_3 + \mathrm{CN} \to \mathrm{HCN} + \mathrm{NH}_2$ when 
$E_\mathrm{t}$ is greater than 4000~K.  Below this energy, the main route of 
production of HCN is the reaction $\mathrm{H} + \mathrm{H}_2\mathrm{CN} \to 
\mathrm{HCN} + \mathrm{H}_2$. 
\end{enumerate}
For CS and H$_2$S, freeze-out is only prevented if $E_\mathrm{t}$ is greater 
than 2000~K.  This is because at this energy both CS and H$_2$S can be directly 
desorbed.  For CS if $E_\mathrm{t}$ is increased further from 2000~K to 3000~K, 
the equilibrium abundance of CS increases again by approximately an order of 
magnitude.  This is due to the direct desorption of H$_2$CS at 2250~K, which 
can increase the CS abundance through the reaction $\mathrm{C}^+ + 
\mathrm{H}_2\mathrm{CS} \to \mathrm{CS} + \mathrm{CH}_2^+$. 

\section{Comparison with existing models and formulations}
\label{sec:comp_models}

\begin{figure*}
\includegraphics{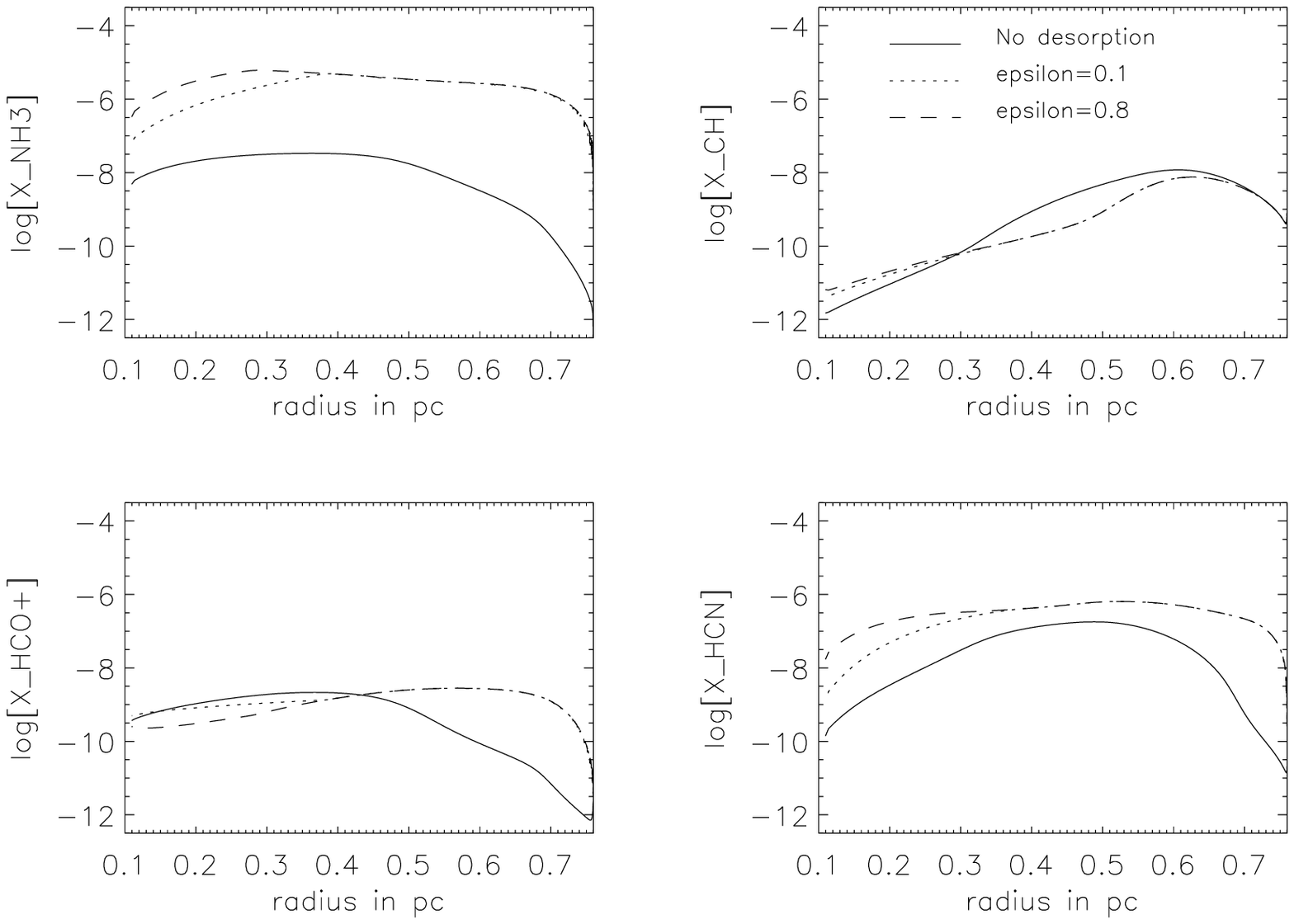}
\caption{The variation of NH$_3$, CH, HCO$^+$ and HCN with radius for a cloud 
collapsing from $n_\mathrm{H}=3 \times 10^3$ cm$^{-3}$ to $10^6$ cm$^{-3}$ at 
10~K including non-selective desorption resulting from H$_2$ formation, as in 
Figure 1 of \citet{b33}}.
\label{fig:W94fig1}
\end{figure*}

We first of all compare our results to those of WRW94, which is the closest
previous study to our own,
then we consider two other studies, \citet{WM98} and \citet{b16}
(hereafter WM98 and HW90 respectively), whose conclusions differed
to ours.  We also discuss various assumptions that are made in most chemical 
models of molecular clouds in which desorption is included, with particular
reference to the formulation of desorption by direct cosmic ray heating by 
HH93.

\subsection{Comparison to WRW94}
WRW94 investigated each of the desorption mechanisms discussed above, in
the case of a collapsing cloud.
They concluded that the only significant mechanism was desorption
arising from H$_2$ formation on grains if it was assumed to be non-selective,
using values for $\varepsilon$ of 0.1 and 0.8.  Since we have found that, for a
static cloud, all desorption mechanisms are significant and can prevent
complete freeze-out, it is therefore worth checking that our model agrees with
that of WRW94 when it is run under similar conditions.

To simulate a collapse, as in WRW94, we have used a spherically symmetric,
isothermal model, undergoing a free-fall collapse.
The collapse is modified by a retardation factor $B$
due to magnetic fields and other factors \citep{b24}.
WRW94 used $B=0.7$, a temperature of 10~K, an initial
density, $n_0$, of $3 \times 10^3$ cm$^{-3}$ and a final density of $10^6$
cm$^{-3}$.  The initial visual extinction of the cloud was 4.4 magnitudes.
Figure~\ref{fig:W94fig1} shows the results from our model when it was run with
these parameters.  We have plotted the abundances of the same molecules that
appear in Figure 1 of WRW94, and have converted time to radius as they did to
allow for easy comparison between our results and theirs.  Even though the
chemical reaction network in our model is much more complicated than in the
WRW94 model, both models agree that desorption has the greatest effect on
NH$_3$ and HCN.  Also, under these conditions, we agree that the other
mechanisms (direct cosmic ray heating and cosmic ray photodesorption) do not
significantly affect the abundances of these molecules.

\subsection{Comparison to WM98}
WM98 carried out an investigation in which they tested the three
desorption mechanisms used in our model, as well desorption by the explosive
chemical reactions of radicals.
Their model used similar physical conditions to ours, except that they adopted
a lower density ($n_\mathrm{H}=2\times 10^4$~cm $^{-3}$).
They found that models that include desorption by H$_2$ formation
(with values of $\varepsilon$ ranging from 0.1 to 1.0)
gave the best agreement to the observed gas phase abundances in TMC-1.

This actually agrees well with our model, since we also find that desorption
by H$_2$ formation with $\varepsilon = 0.1$ is the most efficient mechanism.
WM98 concluded, however, that desorption by direct cosmic ray heating and
cosmic ray photodesorption were too inefficient because they generally
under-predicted the gas phase abundances. Thus, for example, CO, was found to
be underabundant
by approximately two orders of magnitude at an age of 5~Myr using the cosmic
ray induced mechanisms.  We agree that, when comparing the desorption
mechanisms using our standard values of $\varepsilon$, $\phi$ and $Y$,
desorption by H$_2$ formation can indeed maintain gas-phase abundances up to
two orders of magnitude higher than the other mechanisms. But, since these
parameters are so uncertain, we believe it is too early to confidently rule
out cosmic ray heating and cosmic ray photodesorption as important desorption
mechanisms.

\subsection{Comparison to HW90}
In their study, \citet{b16} carried out a purely theoretical analysis of cosmic
ray induced desorption.
They argued that direct cosmic ray heating would not be able to maintain
significant gas-phase CO abundances because the desorbed CO would
be destroyed by reactions with He$^+$ (leading to the formation
of hydrocarbons).  Their analytical calculation of the gas-phase CO
abundance at high densities gives the result
\begin{equation}
n_g(\mathrm{CO}) \propto e^{-\beta t},
\end{equation}
where $\beta$ depends on the rates of the reactions

\begin{equation}
\mathrm{He}^+ + \mathrm{CO} \to \mathrm{C}^+ + \mathrm{O} + \mathrm{He},
\label{eqn:co1}
\end{equation}

\begin{equation}
\mathrm{CO}(\mathrm{g}) \to \mathrm{CO}(\mathrm{s}),
\label{eqn:co2}
\end{equation}

\begin{equation}
\mathrm{CO}(\mathrm{s}) \to \mathrm{CO}(\mathrm{g}),
\label{eqn:co3}
\end{equation}

\begin{equation}
\mathrm{He} + \mathrm{CR} \to \mathrm{He}^+ + \mathrm{e} + \mathrm{CR}
\label{eqn:he1}
\end{equation}
and
\begin{equation}
\mathrm{He}^+ + \mathrm{H}_2 \to \mathrm{H}^+ + \mathrm{H} + \mathrm{He}.
\label{eqn:he2}
\end{equation}

Reactions~(\ref{eqn:co1}) to (\ref{eqn:co3}) are the main formation and
destruction routes for CO, with reactions~(\ref{eqn:co1}) and (\ref{eqn:co3})
referring to the freeze-out and desorption of CO.  Reactions~(\ref{eqn:he1})
and (\ref{eqn:he2}) are assumed to be the main formation and destruction
routes of He$^+$ at high densities.

\citet{b16} calculated $\beta$ to be
\begin{equation}
\beta^{-1} \approx 3 \times 10^6 \mathrm{yr} (n_\mathrm{H}/10^6
\mathrm{cm}^{-3}).
\end{equation}

However, when we redo the calculation with the (updated) rate coefficients for
these reactions used in our model, we obtain values of $\beta^{-1}$ that
are a factor almost six orders of magnitude larger. This implies that
significant gas-phase CO abundances can be maintained over the lifetime of a
typical cloud. The difference arises because we have used a much smaller
desorption rate for CO (given by equation~(\ref{eqn:cr})). So, in fact, using a
{\em smaller} desorption rate allows for a more controlled release of CO into
the gas phase and thus enables the CO to remain in the gas-phase for a longer
period of time.

\citet{b16} also concluded that cosmic ray induced photodesorption would only
be effective in regions where the number density is 10$^3$~cm$^{-3}$ or lower.
They calculated the ratio at which heavy molecules desorb by cosmic ray
induced photodesorption to the rate at which they freeze-out to be
\begin{equation}
R \approx \Big( \frac{X_Z}{10^{-7}}\Big)^{-1} \Big(\frac{n_\mathrm{H}}{10^5
\mathrm{cm}^{-3}}\Big)^{-1} \Big( \frac{\zeta}{10^{-17}\mathrm{s}^{-1}}\Big),
\end{equation}
where $X_Z$ is the fractional abundance of heavy molecules in the gas phase
and $\zeta$ is the cosmic ray ionisation rate. This expression implies
that cosmic ray photodesorption will be more efficient at low densities. The
rates we have used are consistent with this ratio, but our results clearly
indicate that the cosmic ray induced photodesorption
mechanism can still maintain significant gas phase abundances of heavy
molecules at densities as high as $10^5$~cm$^{-3}$, even though the efficiency
of the process is not at its peak at this density.

\subsection{Discussion of other formulations and assumptions}
\label{sec:HH93}
As mentioned earlier, other previous models that have included desorption in
dark clouds tend to only include desorption by direct cosmic ray heating.  One
of the most commonly used formulations is that of HH93, who assumed that
classical ($\sim 0.1~\mu$m) grains are impulsively heated by relativistic
nuclei with energies 20-70~MeV, depositing an average of 0.4~MeV per impact,
raising the local (hot spot) temperature of the grain to 70~K.
The subsequent (thermal) desorption rate is then calculated from the binding
energies of the various molecular species and the cooling profile of the
dust grain.
In practice, the details of the cooling are simplified to a `duty cycle'
approach, so the desorption rate is proportional to the fraction of time spent
by the grains in the vicinity of 70~K,
$f(70~\mathrm{K}) \sim 3.16 \times 10^{-19}$.
The rate coefficient is then given by:
\begin{equation}
\begin{array}{ll}
k_\mathrm{cr-HH93}(i) = f(70~\mathrm{K}) \nu_0(i)
\exp [-E_\mathrm{a}(i)/70~\mathrm{K}]  & \mbox{s$^{-1}$}
\end{array}
\end{equation}
where $\nu_0(i)$ and $E_\mathrm{a}(i)$ are the characteristic adsorbate
vibrational frequency and adsorption energy for species $i$ respectively.

\citet{b11} formulated a similar (but simpler) rate for cosmic ray desorption
(in cm$^{-3}$s$^{-1}$) given by:
\begin{equation}
R_\mathrm{crd-FPW05} = M_s(i) \langle \pi a_\mathrm{g}^2 n_\mathrm{g} \rangle
\gamma \exp \bigg[ \frac{-(E_\mathrm{a}(i) -
E_\mathrm{a}(\mathrm{CO}))}{70~\mathrm{K}} \bigg]
\end{equation}
where $\gamma = 70$~cm$^{-2}$ s$^{-1}$ gives the desorption rate of CO per unit
area of dust grains as derived by \citet{b18}.  This factor takes into account
the cosmic ray flux, $F_{\mathrm{cr}}$, and the efficiency parameter, $\phi$,
which were explicitly included in our rate.  The above rate includes the same
exponential factor as in HH93, but does not take into account variations in
$\nu_0$ for each species.

We decided not to use the above formulations because, bearing in mind
the exponential sensitivity of the desorption rate to the ratio of the hot spot
temperature to the binding energy, the resulting rates are extremely uncertain
and strongly dependent on a number of very poorly constrained free parameters
including the grain size and morphology, the molecular binding energies, the
rate and energy/mass spectrum of the cosmic rays.

We can examine a few of these assumptions in a little more detail:

\begin{enumerate}

\item The formulations given above assume that interstellar dust grains can be
thought
of as homogeneous, symmetrical (spherical, spheroidal, or cylindrical)
entities. In reality, we know that the grains may have very complex
morphologies (`fluffy aggregates') with non-uniform thermal properties and
composition. We can speculate that these grains therefore consist of poorly
thermally connected sub-units, so that both the spot-heating and
the whole grain heating rates may be quite different to what is derived for
spherical, uniform dust grains.

\item Most of these studies also only consider the gas-grain interactions with
large ($\sim 0.1~\mu$m), classical grains. The justification given for this is
that the equilibrium temperature of the (cosmic ray heated) smaller grains is
too high to allow gas-phase species to freeze-out. However, this
differentiation is not apparent in \citet{b18} and there is no direct
observational evidence of the universal presence of
a population of warm dust grains. The smaller grains present a
much larger surface area per unit volume of gas than the classical grains, so
this is an important issue.

\item Concerning the desorption process itself, the assumption is usually made
that the mantle ice is pure, so that the adopted binding/desorption energies
correspond to the pure substance.
Real ices are likely to have complex compositions and
temperature-dependent morphologies. Laboratory work \citep{coll,b5}
shows that the desorption characteristics are drastically modified in such
circumstances in a way that is very species-dependent. Thus, for example, in a
mixed CO/H$_2$O ice, CO is desorbed in
four distinct temperature bands (with $T \sim 25-100$~K).

\item \citet{bri} performed detailed molecular dynamics calculations of
the thermal evaporation process driven by a cylindrical heat pulse,
concentrating on a classical grain of $\sim0.1~\mu$m, heated by a heavy cosmic
ray ion.
Even with the assumption of a single desorption energy/threshold for each
species they still obtained very different results to HH93 -- most notably the
derived H$_2$O and CO desorption rates were found to be an order of magnitude
higher and lower, respectively, as compared to HH93.
However, even here, the calculations make very general assumptions about
the cosmic ray energy spectrum and adopt a spherically symmetric grain with
uniform thermal properties.

\item The issue of the surface chemistry is often not considered
in many of the chemical models which use these desorption formulations. This
would only be valid if one makes the seemingly unreasonable assumption that
the surface residence timescale of adsorbed species is less than the surface
migration timescale for reactive species such as hydrogen atoms.

\end{enumerate}

Because of these various uncertainties we must conclude that although
approximate quantifications of the rate of desorption driven by cosmic ray
grain heating are useful, the numerical details are very poorly constrained and
anything more complicated or specific than our simple approach is very hard to
justify.

The most important conclusion of this section, therefore, is that due to the
considerable uncertainties in the microphysics of ice desorption which are,
as yet, very poorly constrained by laboratory and/or theoretical studies,
it is very difficult to make {\em qualitative} distinctions between different
desorption mechanisms: Each of the desorption models discussed above has its
own merits, but that there is insufficient information to discriminate or
validate these models.
However, we also make the point that, through empirically deduced values of
the depletions, we {\em can} make well-constrained {\em quantitative} estimates
of the desorption efficiencies.
In the next section, we compare our models to observational results. Thus, by
consideration of the deduced desorption efficiencies we are able to comment on
the allowed values of the free parameter in each of the desorption models
($\varepsilon$, $\phi$ or $Y$) that are consistent with the observations.

\section{Comparison with observations}
\label{sec:comp_obsns}

As a consequence of our poor understanding of the theory of ice desorption,
it is proposed that, rather than using desorption efficiencies as an input
to the chemical models we invert the process and use whatever observations we
can safely interpret to constrain, empirically, the nature and efficiencies of
the desorption processes. We can then use that information to predict the
chemical behaviours of other species and generate physical diagnostic
indicators of the molecular clouds.

Observations of CO isotopomers in several pre-stellar cores provide strong
evidence for the depletion of CO onto grains. For example, \citet{obs} were
able to estimate that at least 90~\% of CO is depleted within 5000~AU of the
centre of the pre-stellar core L1689B, by comparing their observations of the
C$^{17}$O $J = 2 \to 1$ line with models which included CO depletion in the
centre of the core.  Since L1689B is a typical pre-stellar core with a
density of $n_\mathrm{H}=1.2 - 1.4 \times 10^5$ cm$^{-3}$ \citep{bac}, it is an
ideal candidate with which we can compare our results.

CO depletion has also been estimated in several other pre-stellar cores, such
as L1544, L1709A, L310, L328, L429 and Oph D \citep{CO}.  In these cases, the
CO depletion is estimated by comparing the observed ratio, $X$, of the
C$^{17}$O and H$_2$ column densities, to the `canonical' abundance determined
by \citet{fre} towards dark cores, $X_\mathrm{can} = 4.8 \times 10^{-8}$.  The
CO depletion factor, $f$, is defined by $X_\mathrm{can}/X$, and was found to
vary from 4.5 (in L1689B, which appears to be slightly less than the depletion
estimated by \citet{obs}) to 15.5 in L429.  Since $X_\mathrm{can}$ is the value
of $X$ obtained in undepleted conditions,
and assuming that there is no selective depletion of different isotopomers of
CO (so the fraction of C$^{17}$O depleted onto grains is the same as the
fraction of total CO depleted), then the fraction of CO depleted onto grains is
given by $1 - 1/f$.
This implies that CO depletion ranges from 78\% to 94\% in these cores.  The
densities of these cores are all of order $n(\mathrm{H}_2)=10^5$ cm$^{-3}$ so it is also
worth comparing these depletion fractions to our model.

Although our results have indicated that including desorption inhibits full
freeze-out allowing abundances to reach an equilibrium, the percentage of CO
frozen-out onto grains is actually greater than 98\% ($f=50$) for all of the
mechanisms investigated, which is consistent with the observations above.

If we assume that L1689B is in a state of equilibrium, the values of
$\varepsilon$, $\phi$ and $Y$ needed to give the 90\% freeze-out estimated in
L1689B are approximately 0.5, $1.3 \times 10^7$ and 5.5 respectively.
We calculated these values by assuming that freeze-out and desorption of CO
are the main processes governing the CO abundance, so by equating these rates
with 90\% of CO on the grains and 10\% remaining in the gas phase we can
predict the efficiencies needed.  We then ran the model with these values of
$\varepsilon$, $\phi$ and $Y$ to confirm that the observed
CO depletion is obtained.
Since it is possible that L1689B is not in equilibrium and further freeze-out
could be achieved, this gives us an upper limit for these parameters.

If we use the values estimated by \citet{CO}, for L1689B (in which they deduce
that CO is depleted by 78\%) we obtain even higher limits for
$\varepsilon$, $\phi$ and $Y$ of 1.4, $3.3 \times 10^7$ and 14.0
respectively.  For L429, in which a CO depletion of 94\% was deduced, the
values for $\varepsilon$, $\phi$ and $Y$ are reduced to 0.31,
$7.5 \times 10^6$ and 3.2 respectively.  The values for all the other cores
studied by \citet{CO} would lie in between the values for L1689B and L429.

These values are much higher than the estimates used in our model, indicating
that if these desorption mechanisms operate then they are very efficient.
However, using these observations it is not possible to determine which of the
mechanisms is operating, or if it is a combination of all three.

\section{Conclusions}
\label{sec:conclusions}

There are two major conclusions from this study: Firstly, we have shown that
the usual assumption that cosmic ray desorption is the the most effective
desorption mechanism in dark molecular clouds is not always valid.
All three desorption mechanisms that we have considered (desorption via H$_2$
formation on grain surfaces, direct cosmic ray heating and cosmic ray
photodesorption) have been shown to
have significant effects on the gas phase abundances in quiescent dark
molecular clouds and so should not be neglected in chemical models.
These processes all operate on timescales of the order of $\sim 10^6$ years.
Each of the processes is capable of preventing total freeze-out, but in an
equilibrium quiescent dark cloud of density 10$^5$ cm$^{-3}$, the predicted
percentage of freeze-out is always greater than 98\%. This figure is in good
agreement with the observations (e.g. L1689B).

Addressing the specifics of the desorption processes,
desorption via  H$_2$ formation, if it is efficient ($\varepsilon = 0.1$), is
the most effective mechanism. However the complete freeze-out of
some species, such as CS and H$_2$S, can only be prevented by the
cosmic ray photodesorption mechanism, which is non-selective.
The relative importance of the three mechanisms appear to be
insensitive to variations in the density and the initial atomic to molecular
hydrogen ratio.
For the selective desorption mechanisms, choosing the threshold adsorption
energy, $E_\mathrm{t}$, (such that molecules with adsorption energies less than
$E_\mathrm{t}$ will be desorbed), can have a strong effect on the chemistry,
particularly on molecules such as NH$_3$, HCN, CS and H$_2$S.

Secondly, our understandings of the chemical and physical structure of dust
grains and the physical processes which drive desorption are, as yet,
very incomplete and a purely theoretical approach to the problem is inadvisable.
In this study we have investigated three particular desorption mechanisms
(desorption via H$_2$ formation, direct cosmic ray heating and cosmic ray
photodesorption) and have used the observed molecular depletions to
constrain the poorly-determined free parameters in desorption
processes. This is the first attempt at an empirical determination of the
desorption efficiencies. Bearing in mind the huge uncertainties
in these efficiencies, this empirical approach is the one that
we recommend for use in future studies of interstellar chemistry where
gas-grain interactions play an important role.

\section*{Acknowledgements}
JFR is supported by a PPARC studentship.

%
% References
%

\bsp

\label{lastpage}

\end{document}